\documentclass{iopart}
\usepackage{graphicx}

\begin{document}
\title[]{Nanoroughened plasmonic films for enhanced biosensing detection}

\author{Eric Le Moal$^1$\footnote{Present address:
Institut f\"{u}r Physikalische und Theoretische Chemie, Universit\"{a}t Bonn, Wegelerstrasse 12, 53115 Bonn, Germany}, Sandrine L\'{e}v\^{e}que-Fort$^2$,
Marie-Claude Potier$^3$ and Emmanuel Fort$^3$}

\address{$^1$ Institut Langevin, ESPCI ParisTech, CNRS UMR 7587, Universit\'{e} Paris Diderot, 10 rue Vauquelin, 75231 Paris Cedex 05, France}
\address{$^2$ Laboratoire de Photophysique Mol\'{e}culaire, Universit\'{e} Paris-Sud, CNRS UPR 3361, 91405 Orsay cedex, France}
\address{$^3$ Laboratoire de Neurobiologie et Diversit\'{e} Cellulaire, ESPCI ParisTech, CNRS UMR 7637, 10 rue Vauquelin, 75231 Paris Cedex 05, France}
\ead{emmanuel.fort@espci.fr}

\begin{abstract}
Although fluorescence is the prevailing labeling technique in biosensing applications, sensitivity improvement is still a striving challenge. We show that coating standard microscope slides with nanoroughened silver films provides a high fluorescence signal enhancement due to plasmonic interactions. As a proof of concept, we applied these films with tailored plasmonic properties to DNA microarrays. Using common optical scanning devices, we achieved signal amplifications by more than 40-fold.
\end{abstract}
%\submitto{\NT}
\maketitle

\section*{Introduction}
Most biosensing applications share a crucial need for high detection sensitivity. Because of its unmatched sensitivity, fluorescence has become the prevailing labeling technique. Improvements in detection sensitivity is the subject of intensive research, in particular through the development of novel fluorescent probes~\cite{Bruchez1998,Medintz2005} and smart detection methods~\cite{Hell1992,Oheim1999}.

In this context, mirror substrates appeared in the market to improve the fluorescence signal in biosensing applications involving classical epi-illumination. Their ability to enhance fluorescence resides in their reflection properties, obtained by coating the surface of cleaned glass slides either with stacked dielectric layers~\cite{Choumane2005} (multilayer Bragg mirror) or a continuous metallic film (e.g., silver or gold). The presence of a reflective surface coating modifies both excitation and emission processes, and additionally reshapes fluorescence emission lobes~\cite{LeMoal2007,LeMoal2007b}.

Although easier and cheaper to manufacture, silver and gold mirrors have a major drawback over dielectric mirrors. Fluorescence is partly quenched by the presence of the metal in its vicinity. In particular, the fluorophores couple to non-radiative surface plasmon (SP) modes of the metallic thin-film through their near-field components~\cite{Barnes2003,Fort2008}. SPs are transverse collective oscillation modes of the conduction electrons that propagate along the surface of a conductor. They are associated to evanescent electromagnetic (EM) fields which exponentially decay in intensity with increasing surface distance~\cite{Raether1988,Barnes2003}. The unique properties of SPs fostered numerous technological breakthroughs in optics, optoelectronics~\cite{Ozbay2006,Barnes2003} and nanosensors~\cite{Anker2008}. In the present work, we applied these concepts to biosensing devices. Our approach is characterized by its simplicity of use and its ability to be directly operated with standard optical setups including epi-fluorescence microscopes and scanning devices.

\begin{figure}
\includegraphics[width=12cm]{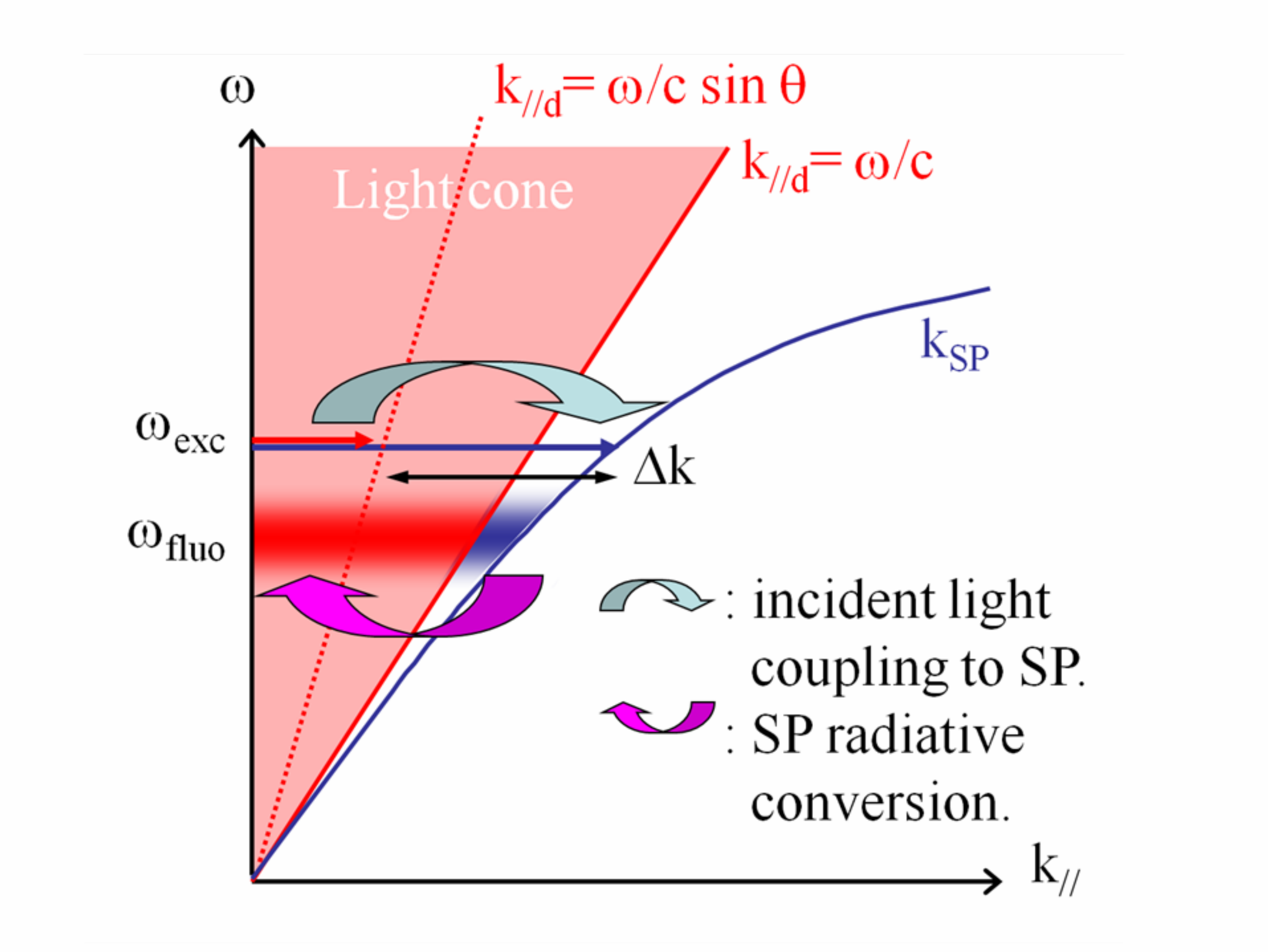}
\caption{Dispersion curves (pulsation $\omega$ versus in-plane wavevector $k_{\parallel}$) of a free-space photon (\emph{red lines}) and a SP mode at a plane metal surface (\emph{blue line}). The arrows indicate the coupling between light and SP resulting from film roughness which provides the needed missing momentum $\Delta k$. $\omega_{exc}$ and $\omega_{fluo}$ stand for the pulsations of the excitation and emission fields, respectively. $\theta$ is the incidence angle of the excitation field and is defined with respect to the normal direction to the metal surface.} \label{fig:dispersion}
\end{figure}

The main objective was to turn the unfavorable energy transfer to SPs into a positive radiative process, to boost the fluorescence amplification. This requires the SPs to be converted into light, which is precluded on smooth metal surfaces, due to the momentum conservation principle~\cite{Fort2008}. Figure~\ref{fig:dispersion} shows the SP dispersion curve and the light cone associated to the dispersion curves of propagative photons. At any given frequency, SP wavevector is larger than that of free-space photons. Nevertheless, the SP energy can be recovered by corrugating the metal
surface and exploiting SP scattering effects. In that case, the difference in wavevector $\Delta k$ can be provided by the Bragg vectors associated with the
spatial periodicity of the metallic surface~\cite{Raether1988}.

In our experiment, we use randomly roughened surfaces which may be considered as multicorrugated surfaces and seen as a weighted sum of periodically corrugated surfaces. Whereas the use of a sinusoidal grating to couple SP to photonic modes requires lithographic procedures and specially designed observation geometries~\cite{Sullivan1994}, roughened surfaces can directly be made by thermal evaporation and be used on standard epifluorescence microscopes and scanning devices (e.g., DNA chip readers).

\section{Methods: DNA microarray preparation}
Modification of glass substrates with silane adhesion was achieved by immersion of glass slides in a methanolic solution containing 2\% solution of (3-mercaptopropyl)-trimethoxysilane (Sigma Aldrich), 1\% acetic acid, 4\% MilliQ-filtered water during 2 hours at room temperature and protected from ambient light. This step was performed immediately after the deposition of the amorphous alumina thin film. Silanized glass slides were then rinsed in HPLC grade methanol and were gently agitated for 5 min to remove uncoupled reagent. The silanized glass slides were dried by centrifugation during 3 min at 500~rpm. The absorbed silane layer was cured at 100$^\circ$C for two hours. Silanized glass slides were stored in a clean and dust free box at room temperature prior further use.

A solution of heterobifunctionnal PEG (N-hydroxysuccinimide-poly\-ethylene\-glycol-vinyl\-sulfone, NHS-PEG-VS, MW 3.4kDa) was prepared by dissolution of 3.75 mg in 1~ml of PBS 1x (phosphate buffer saline) pH 7.4. 50 $\mu$l of this solution was spread on one side of the silanized glass slide in order to form a liquid film on the slide and is left during 45 min at room temperature. The PEG
functionalized supports were then rinsed twice in MilliQ-filtered water and dried under an argon stream. Prior to spotting, the peggylated glass slides were stored safe from dust and light.

Prior to spotting, the lyophilized 50mer oligonucleotide probe 5'NH2-AGACT\-TCTCCAGG\-TGATTGTCA\-AAGGAG\-ATCTGT\-CTGAGGGG\-AGGTAAAA corres\-ponding to the GABA$\alpha$ 6 gene (NM\_008068) present on the Neurotrans Platform GPL4746 at www.ncbi.nlm.nih.gov/geo was resuspended in 150 mM phosphate buffer pH 8.3 at a 10 µM concentration. The oligonucleotide was applied onto the slides by contact printing using a Biorobotics Microgrid II instrument (Genomic solutions, Ann Arbor, Michigan) with one BioRobotics MicroSpot 2500~pin. Oligonucleotide spots were 80-110~$\mu$m in diameter, and the center-to-center between two adjacent spots was 300~$\mu$m. After spotting, the arrays were kept in a humidity chamber with 60\% relative humidity for a time period between 10 and 24~h. Excess of amine-reactive groups were deactivated by keeping the slides for 30~min at 50$^\circ$C in a solution containing 50~mM ethanolamine, 0.5~M Tris-HCl, pH~9.0. The slides were washed twice with MilliQ-filtered water and incubated for 30~mn in a solution 4X SSC, 0.1\%. Then, the slides were washed twice with MilliQ-filtered water and finally dried by centrifugation during 3 min at 500~rpm.

The target was a 15mer 5'-labeled Cy5 oligonucleotide Cy5-TTTT\-ACCTCC\-CCTCAGACA\-GATCTCC\-TTTGACAA\-TCACCTGG\-AGAAGTCT complementary to the 50mer probe. Prehybridization of the microarray was
carried out with 20~ml solution containing 18~ml Dig Easy buffer (Roche Diagnostic) and 2~ml of denatured salmon sperm DNA over 1~h at 42$^\circ$C. The slide was washed with 2X SSC and then hybridized at 42$^\circ$C for $12\pm 14$~h in a similar solution to which was added $\mu$M of the oligonucleotide target. Hybridization was carried out in a coverslip hybridization cassette (Corning Inc), which was kept humid by addition of 20 ml water in the chamber. After hybridization, slides were washed for 2 min in 2X SSC/0.2\% (v/v) SDS, for 2 min in 0.2 SSC/0.2\% (v/v) SDS and for 2 min in 0.2X SSC, and then dried under a stream of nitrogen. Detection of the fluorescence signals was done using a laser scanner (GenePix 4000A from Axon Instrument, CA) and analyzed with the GenePix 3.01 software.

\section{Sample preparation and characterization}
To grow randomly nanoroughened silver films of precisely controlled roughness, we developed a two-step deposition technique which basically consists in ``pulling a blanket over nanopillows". Silver was thermally evaporated onto cleaned standard microscope slides in high vacuum conditions (base pressure $10^{-9}$mbar). First, an islandized silver film (the ``nanopillows") was grown onto the substrate at high temperature. Once the substrate cooled down to room temperature (RT), the islandized film was covered with an additional metal layer (the ``blanket"), to obtain a continuous film.

The roughness of the silver film can be tailored by controlling the experimental conditions, notably the substrate temperature during metal deposition. Heating the substrate can activate the coalescence of the metal islands that form at the first steps of the film growth. This yields larger metal particles and this increases the film roughness. However, raising the substrate temperature also pushes back the percolation threshold, which makes it more difficult to form a continuous film. Consequently, one has to proceed in two steps to control both the roughness of the silver film and the area coverage of the metal on the glass substrate.

\begin{figure}[h!]
\includegraphics[width=12cm]{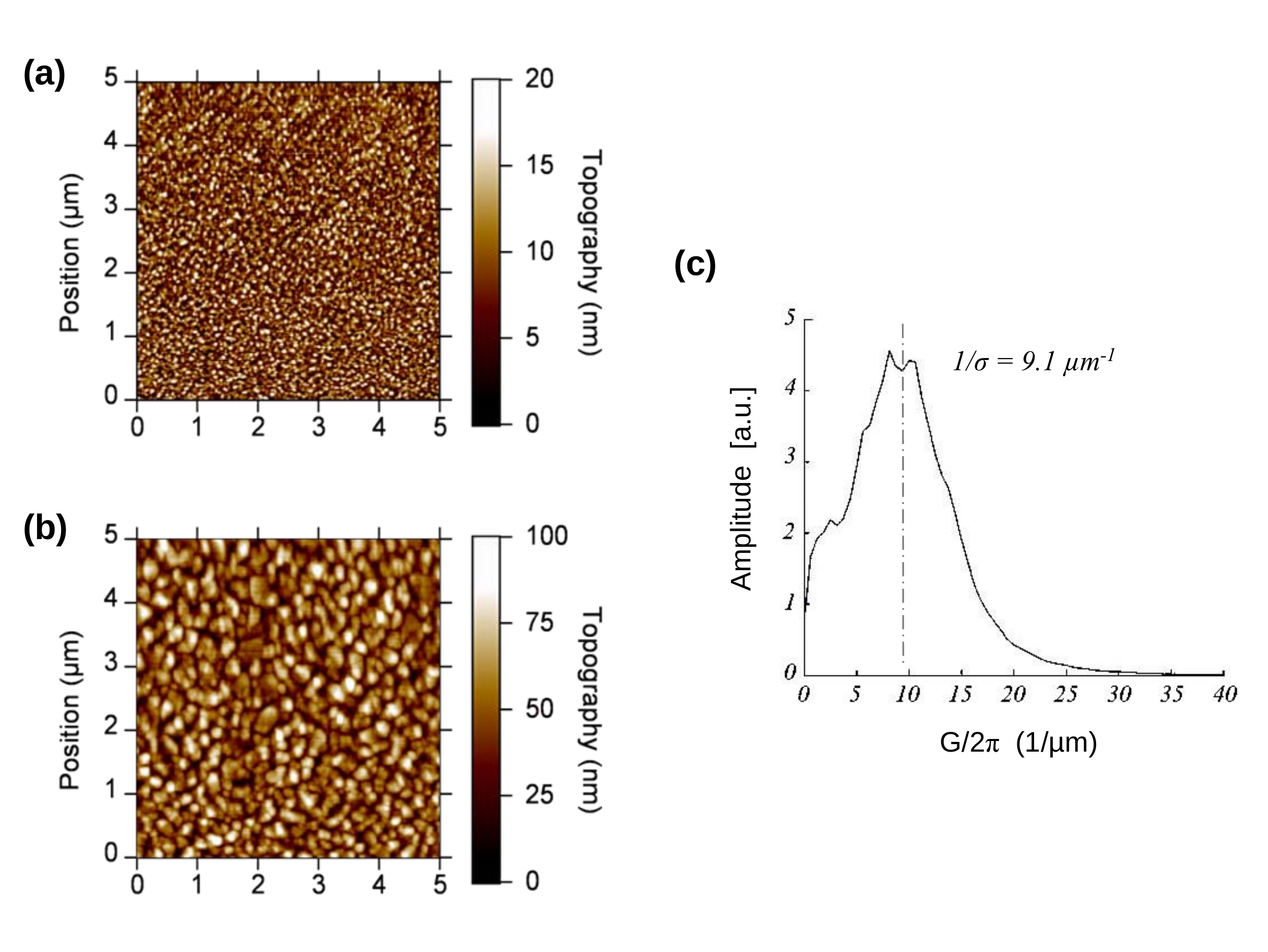}\\
\caption{AFM images representing the surface topographical image (a) of a ``smooth" silver film (\emph{sample}~1) and (b) a nanoroughened silver film (\emph{sample}~5) in a 5~$\mu$m~$\times$~5~$\mu$m area. (c) Spatial frequency distribution of a roughened surface obtained from AFM topography measurements (associated to figure~\ref{fig:topography}b). The two step deposition technique has been used to tune the correlation length to the value of 0.11~$\mu$m.}\label{fig:topography}
\end{figure}

The surface topography of silver-coated substrates is mapped by Atomic Force Microscopy (AFM). The auto-correlation analysis of the AFM images gives two essential parameters to characterize the surface roughness, namely the root-mean-square (rms) roughness amplitude ($\delta$) and the transverse correlation length ($\sigma$). From the analysis of AFM images, we demonstrated that both the root-mean-square roughness amplitude ($\delta$) and the transverse correlation length ($\sigma$) of the nanoroughened film can be gradually increased by tailoring the size and shape of the ``nanopillows". Figure~\ref{fig:topography} shows the topography of a ``smooth" sample (a) and of a nanorough sample (b) measured by the AFM. Figure~\ref{fig:topography}c shows an example of the spatial frequency distribution obtained from image (b). The correlation length $\sigma$ is 0.11~$\mu$m.

The obtained values of the topographic parameters $\delta$ and $\sigma$ for various samples are shown in Table~\ref{tab:roughness}. The roughness analysis of the AFM images revealed a strong morphological modification when silver is deposited on a substrate that has been coated with a silver island film first rather than directly on a bare substrate (\emph{Sample~1}). The roughness parameters $\delta$ and $\sigma$ are respectively increased more than fourfold ($\times 4.6$) and twofold ($\times 2.3$) in \emph{Sample~5} compared to \emph{Sample~1}. The gradual increase of $\delta$ and $\sigma$ with the amount of silver that is deposited at high temperature (see Table~\ref{tab:roughness}) demonstrates the efficiency of the two-step technique to
control the roughness of nanoroughened silver films.

\begin{table}[h]
        \rule{13cm}{1pt}\\
        \begin{tabular}{lcccccc}
        Ref. & \multicolumn{2}{c}{Deposition steps} \\
        \cline{2-3}
        & Ag at 300$^{\circ}$C & Ag at RT & ~~~$\delta$~~~ & ~~~$\sigma$~~~ & TEM Obs. & $G$ \\
        \hline
        \emph{Sample}~1 & 0 & 60 & 3.5 & 47 & continuous & 28\\
        \emph{Sample}~2 & 20 & 0 & 5.5 & 55 & islandized & 17\\
        \emph{Sample}~3 & 20 & 60 & 6.5 & 59 & quasi-continuous & 35\\
        \emph{Sample}~4 & 40 & 60 & 11.5 & 85 & quasi-continuous & 48\\
        \emph{Sample}~5 & 60 & 60 & 16 & 110 & quasi-continuous & 52\\
        \end{tabular}
        \rule{13cm}{1pt}\\
        \bigskip
        \caption{Morphological parameters and fluorescence enhancement factor $G$ associated to various samples. The fluorescence enhancement factor $G$ that was measured on
        \emph{Sample~1-5}, after deposition of a spacer layer of optimal thickness and a low-density deposit of fluorescent dye (Cyanine~3). By \emph{quasi-continuous} films, we refer to
        percolated films that exhibit a metal area coverage $\Theta>0.9$ (measured by TEM).}
        \label{tab:roughness}
\end{table}

\begin{figure}
\includegraphics[width=12cm]{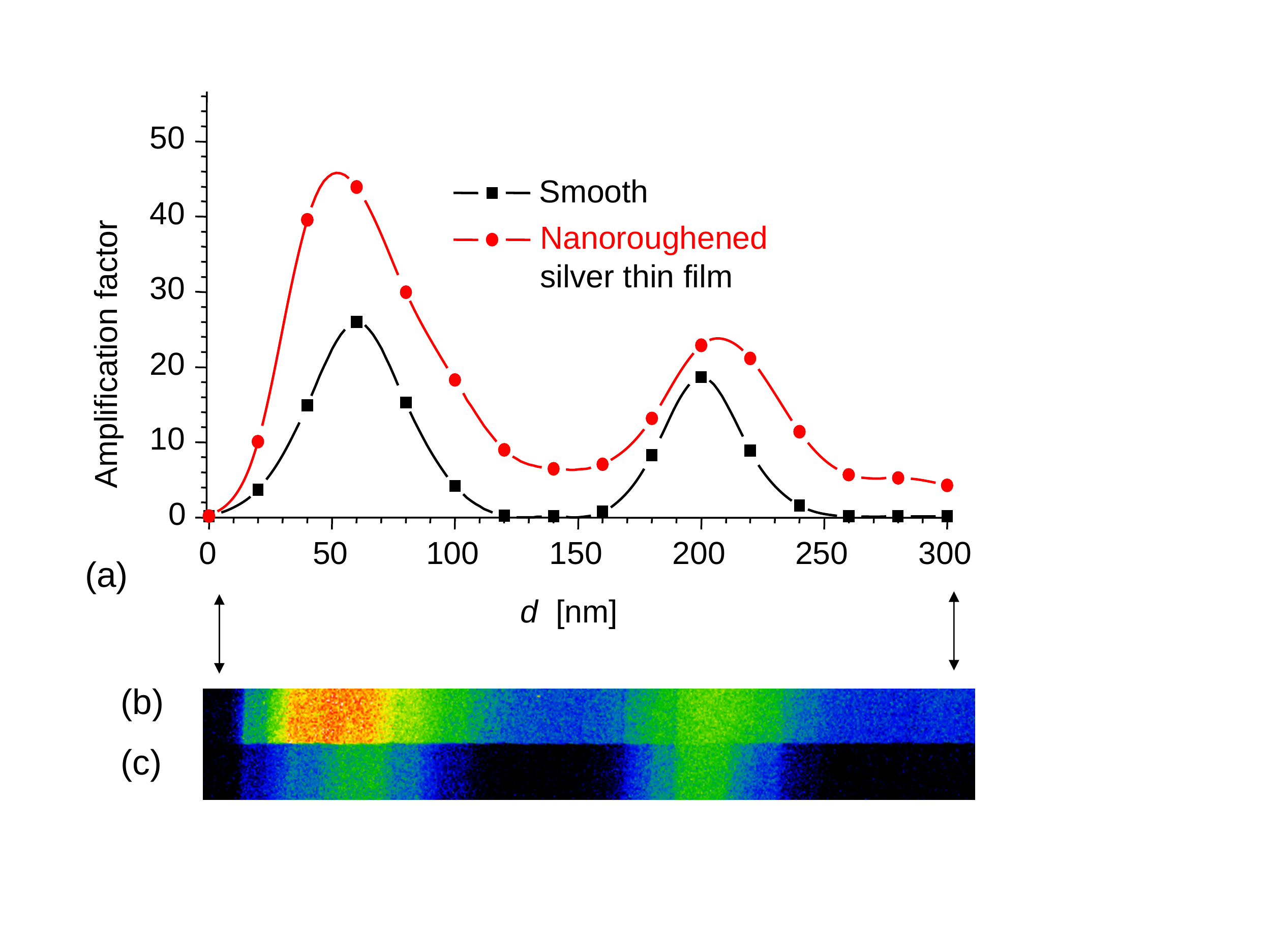}
\caption{(a) Fluorescence enhancement factor versus spacer layer thickness $d$ on a smooth (\emph{sample}~1) and a nanoroughened (\emph{sample}~2) silver films, for thermally evaporated Cyanine~3 dye (excitation wavelength 543~nm). The enhancement factors are obtained directly from the analysis of the fluorescence images of samples with a discrete spacer increase with step thickness of 10~nm, (b) and (c) respectively.} \label{fig:Dye}
\end{figure}

Fluorescent dye molecules (Cyanine~3) were deposited by thermal evaporation onto silver-coated substrates to investigate their optical properties. The distance between the fluorophores and the silver film was tuned by adding a spacer layer of amorphous alumina by pulsed laser deposition technique. Figure~\ref{fig:Dye}a shows the fluorescence enhancement factor as a function of spacer layer thickness on a smooth (black squares - \emph{Sample 1}) and nanoroughened (red circle - \emph{Sample 5}) silver films, for thermally evaporated Cyanine~3 dye upon laser excitation at 543~nm. The enhancement factors are obtained directly from the analysis of the fluorescence images on samples prepared with various thicknesses. Figure~\ref{fig:Dye}b and Figure~\ref{fig:Dye}c show respectively images of sample with increasing spacer layer from 0 to 300 nm with step 10~nm for a smooth and a nanoroughened thin film respectively.

As expected, the emission of fluorophores above a plane (mirror-like) silver film is clearly modulated by the excitation field, due to the superposition of the incident and reflected electromagnetic (EM) waves on the mirror. Signal maxima were observed for fluorophores positioned at antinodes of the excitation field, \emph{i.e.}, for $d\approx \lambda/4+n\lambda/2$ where $n$ is an integer (if neglecting the penetration depth of the EM field in the metal). The two first maxima were found at spacer layer thicknesses of $d_1=60$~nm and $d_2=200$~nm. At $d_1$, fluorescence signal was amplified by a factor $G=26$ on the smooth silver film and $G=45$ on the nanoroughened silver film. Roughness thus induce a 1.7-fold gain in fluorescence detection. In a first order approximation, this signal enhancement can be explained by the radiative conversion of the energy transferred to SPs. While on smooth films this energy only dissipates in the metal and cannot be detected, roughness induces light scattering. A more quantitative evaluation of the enhancement effect due to roughness is very complicated since the measured roughness is sufficient enough to induce perturbations of the SP propagation and to induce SP localization. SP are associated with evanescent EM fields, the spatial extension of which is of about $\lambda/2$, hence roughness effect is mainly limited to the first optimal spacer layer thickness ($d_1=\lambda/4$). It is noteworthy that detected intensity at $d=0$ is almost zero on both smooth and nanoroughened silver films. For separations of < ~5 nm, a very efficient nonradiative energy transfer from the molecule to the metallic surface occurs. It is understood that the near-field of the molecule penetrates the metal and excites electron-hole pairs (excitons) that quench fluorescence. Classically, the process is described by induced charge-density oscillations also called lossy surface waves (LSW), which can be interpreted as dipole-dipole energy transfer to a volume of dipoles below the surface~\cite{Fort2008}. As the distance to the metallic surface increases, the relative proportion of transferred energy to LSW rapidly falls while that to SP increases. Consequently, a spacer layer of optimized thickness is necessary between the fluorophores and the metallic surface not only to maximize fluorophore excitation rate but also to increase the efficiency of SP coupling and SP conversion into light.

On the nanoroughened film, fluorescence enhancement is found to be less spatially selective than on the plane film (on the enhancement-\emph{vs}-distance graph, the first enhancement peak is about 40\%-wider at half height). This can be an advantage for the global amplification factor when using two types of fluorophores, that differ by several tens of nanometers in excitation wavelength. In particular, dual-labeling is commonly used in DNA microarrays to provide a reference signal or to reduce bias (for instance, with Cy3/Cy5 dyes excited respectively at $\lambda_{exc}=550$~nm and $\lambda_{exc}=649$~nm).

The maximum enhancement factor for a given sample at the optimal spacer layer thickness is given in Table~\ref{tab:roughness}. As $\delta$ and $\sigma$ are increased, $G$ gets larger from 28 for \emph{Sample~1} up to 52 for \emph{Sample~5}. We presume that this effect is due to the enhancement of SPP scattering, which results in increased radiative efficiency~\cite{Moreland1982,Worthing2001}. A 52-fold signal enhancement was measured on a randomly nanoroughened silver film with optimized spacer layer thickness ($d=60$~nm) and roughness parameters ($\delta=16$~nm and $\sigma=110$~nm), as compared to standard glass slide (see Figure~\ref{fig:Dye}). This is nearly twice as much as on a smooth silver film in the same conditions.

\begin{figure}[h!]
\includegraphics[width=12cm]{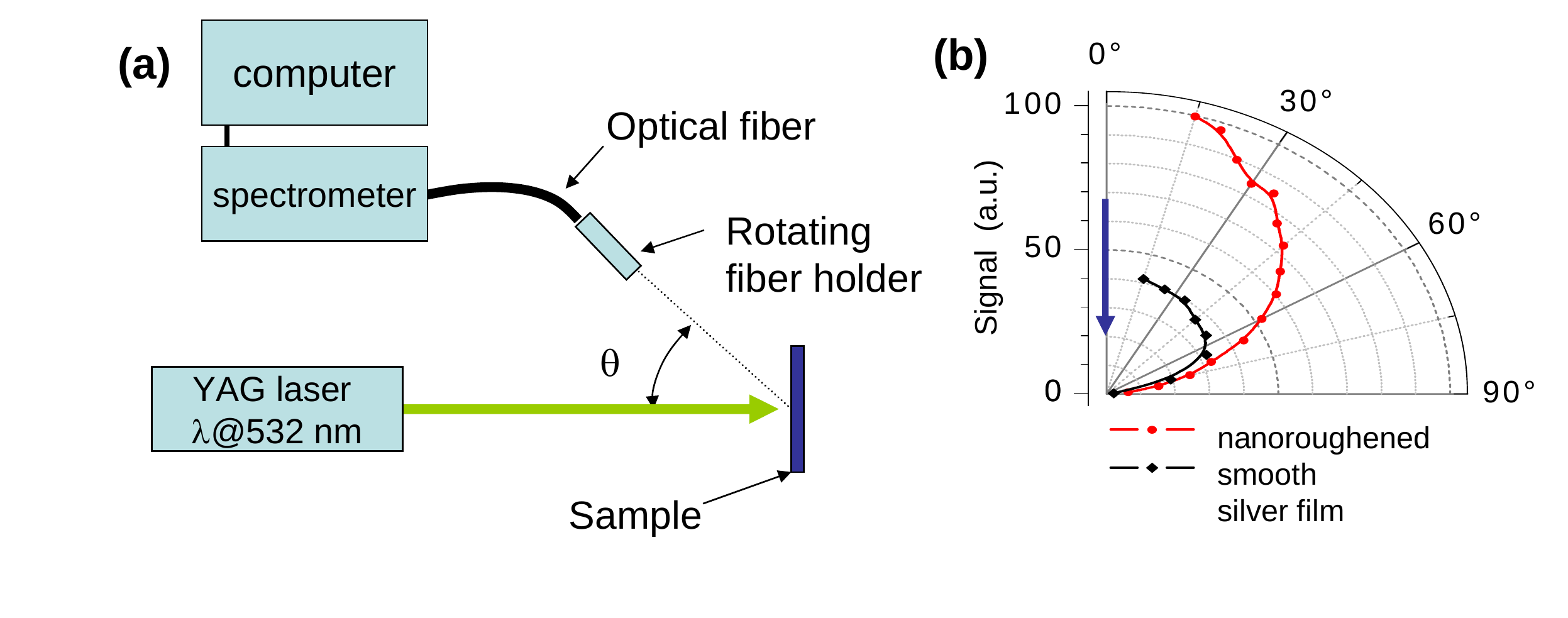}
\caption{(a) Schematics of the experimental set-up was designed to measure the angular distribution of the fluorescence emission; (b) Fluorescence emission lobe measured with a rotating detector. The signal intensity is plotted as a function of the angle to the surface normal. The excitation light, with a wavelength of 532~nm, is orthogonal to the sample surface (indicated by the arrow). The fluorescence signal comes from Rhodamine B fluorophores evaporated on a nanoroughened and on a flat silver film with an alumina spacer layer of 60 nm.}\label{fig:lobes}
\end{figure}

Whereas the monoperiodicity of a sinusoidal grating imposes a very strict condition on photon momentum for coupling to SPs (thus resulting in a highly directional emission)~\cite{Andrew2001}, random roughness advantageously loosens this condition. We designed a simple device to measure the angular distribution of the fluorescence emission. Figure~\ref{fig:lobes}a shows the schematics of the optical setup. An optical fiber is carried by an arm spinning around the sample in order to collect the emitted light in all directions. The incident laser reaches the sample at normal incidence. The excitation source is a continuous Nd:YAG laser at 532~nm, well adapted to Rhodamine~B absorption band. Collected light is acquired using a spectrometer in order to discriminate fluorescence from scattered excitation light and to study spectrum modifications as the angle of collection changes. Figure~\ref{fig:lobes}b shows emission patterns of an evaporated thin layer of Rhodamine B deposited onto a flat and corrugated sample. Emission intensity is enhanced on the corrugated sample by about two fold. Both samples yield to wide and nearly isotropic emission patterns. Due to the broad spatial frequency spectrum, randomly roughened surfaces allow efficient coupling between light and SPs. An important part of SP energy can thus be recovered by re-radiation~\cite{Moreland1982}.

It is noteworthy that random roughness also allows a coupling between SPs and the incident light focused on the metallic surface through a microscope objective (i.e., with a broad distribution of incidence angles). The resulting field associated to SPs is expected to be more intense in the vicinity of the metallic surface than in the case of a smooth metallic surface and to further contribute to signal enhancement.

The surface coverage of the metallic films was calculated from Transmission Electron Microscope (TEM) images (see Tab.~\ref{tab:roughness}). TEM images show that the films are far beyond the percolation threshold and possess only sparse holes. Comparison with the islandized films reveals the importance of the continuous or quasi-continuous nature of the nanoroughened films in the enhancement phenomenon. Although the roughness parameters of \emph{Sample~3} are less than 20\% higher (for $\delta$) and 10\% higher (for $\sigma$) than that of \emph{Sample~2}, the quasi-continuous silver film (\emph{Sample~3}) yields a twofold higher fluorescence enhancement factor than the silver island film (\emph{Sample~2}). This is probably to be related with the spatial distribution of the EM field. In continuous thin films, this distribution is determined not only by the local topography but also by large-scale geometrical structures~\cite{Arya1985,Aussenegg1987}.

Besides, we measured an apparent photostabilization of the fluorophores on nanoroughened slides, which is of practical interest for sensing applications. In the optimal sample enhancement configuration (sample~5 with 60~nm of spacer layer), about 10-fold more emitted photons are detected before fluorophore photobleaching as compared to standard microscope slides. Such a stabilization is also observed on smooth metallic samples. This effect is due to both a reduction of the photobleaching rate and an improvement of the collection efficiency ~\cite{Enderlein1999}. Close to the metal, coupling to SPs provides fluorophores with additional decay routes; hence fluorophore lifetime is shortened. Since the mean number of excitation/emission cycles that fluorophores can achieved before photobleaching is inversely proportional to their lifetime, shorter lifetimes yield lower photobleaching rates. Collection efficiency also benefits from the presence of a metallic film because it can redirect a significant part of the emitted light towards the entrance of the microscope objective.

We, then, used the nanoroughened samples to DNA microarrays as a proof of concept for biosensing applications. We detected DNA hybridization using a single pair of perfectly complementary sequences.

\section{DNA microarray application}
\begin{figure}
\includegraphics[width=12cm]{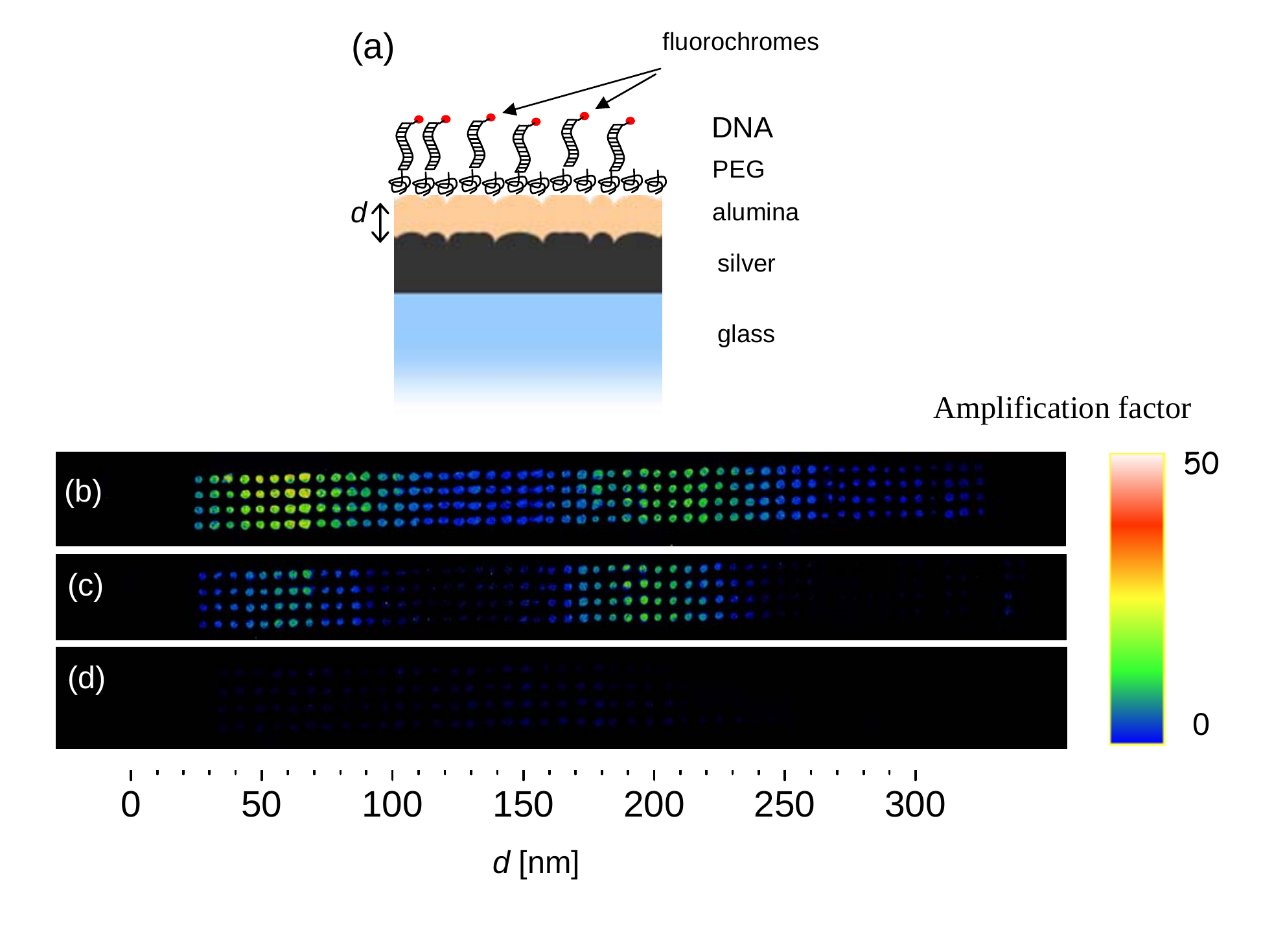}
\caption{(a) Schematic of the fluorescence-enhancing substrate DNA chip. Fluorescence images of rows of DNA spots, on nanoroughened (b) and smooth (c) silver films and on a standard glass substrate (d) with increasing spacer thickness from 0 to 300 nm, with 5 nm steps.} \label{fig:DNA}
\end{figure}

Figure~\ref{fig:DNA}a shows a schematics of the DNA microarray on nanoroughened silver sample. We checked that the nanoroughened silver films were chemically compatible with protocols for covalently binding DNA probes~\cite{Ait2007}. Figure~\ref{fig:DNA} shows the measured fluorescence signal for a (b) nanorough, (c) smooth silver coated substrate and (d) a standard glass slide for spacer layer thickness ranging from 0 to 300 nm with 5 nm steps. The detected fluorescence modulations are very similar to those obtained with evaporated cyanine dyes (see Figure~\ref{fig:Dye}b and~\ref{fig:Dye}c). Maximum 24-fold and 42-fold enhancements were measured on DNA microarrays for a spacer thickness of $d_1=60$~nm on smooth and nanorough silver films. The enhancement is about 20\% lower than for dye layers. However, the enhancement factor on roughened silver films still is 1.7-fold higher than that on smooth ones. Our results also confirm a broader range of spacer thicknesses for fluorescence amplification than on smooth films. The nanoroughened silver films will thus be more suitable for dual-labeling in DNA microarrays than smooth ones.

In DNA microarray applications, the spatial homogeneity of the signal amplification is as important as the gain factor. The radiative conversion of SPs involves complex processes~\cite{Fort2008} and strong roughness notably disrupts SP propagation along the metal interface. SP spatial localization in metallic protrusions induces strong optical resonances which are associated with intense evanescent EM fields (often referred to as ``hot spots")~\cite{Sarychev2000}. This implies a strongly heterogeneous enhancement at the nanometer-scale.

\begin{figure}
\includegraphics[width=12cm]{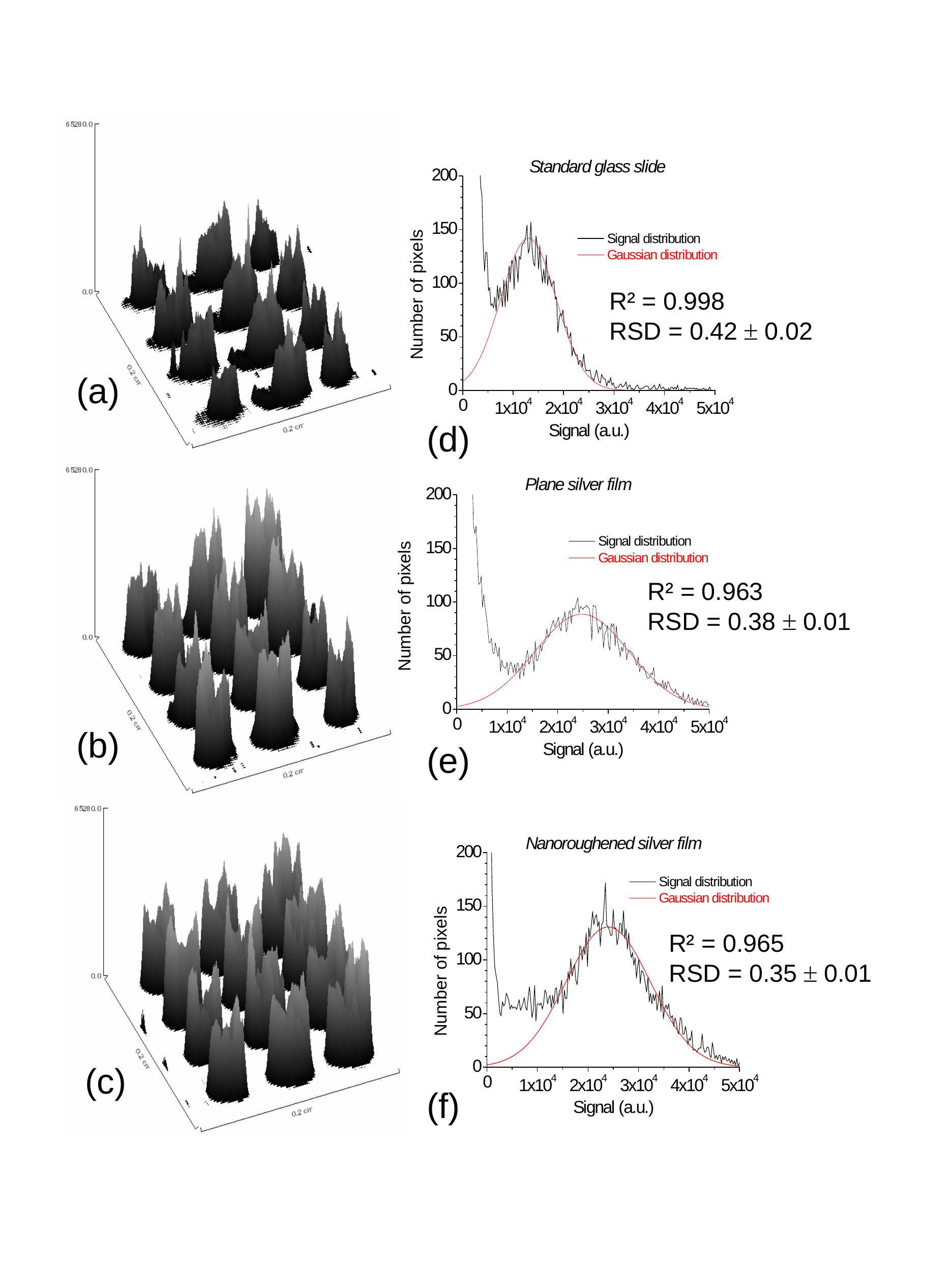}
\caption{Fluorescence signal distribution on a DNA microarray: 3D representations of close ups taken from the fluorescence image of a DNA microarray. The signal intensity, at each pixel of the fluorescence image, is both plotted along the z-axis and coded in levels of grey. The microarray substrates are of three different types: a bare glass slide (a); a glass slide coated with a smooth silver film (b) and with a nanoroughened silver film (c). The imaged regions of interests are 4~mm$^2$ in surface area (i.e., 160000~pixels), each comprising nine similar DNA spots. Associated intensity histograms (d), (e) and (f) show the signal distributions. The experimental data are fitted with a standard Gaussian distribution function (red line). $R^2$ is the coefficient of  determination, $RSD$ is the relative standard deviation.} \label{fig:distribution}
\end{figure}

Figure~\ref{fig:distribution}a-c shows the image of the fluorescence signal obtained with the maximum scanner resolution (5 $\mu$m/pxl) on 9 DNA spots (a) for a standard glass substrate and (b) for smooth and (c) nanorough metallic coated subtrates. Figure~\ref{fig:distribution}d-f shows the associated histogram of the image intensity. The intensity distributions were fitted with a standard Gaussian distribution function (see continuous red line on the histograms). In each case, we calculated the coefficient of determination R$^2$, namely the squared coefficient of correlation between the experimental distribution of the signal and the fitted Gaussian functions. The relative standard deviation RSD corresponds to the distribution width. The relative constancy of both $R^2$ and $RSD$ for each type of substrate indicates that the presence of the metallic thin film does not affect the signal dispersion on the DNA microarray, at the scale of the microarray-reader spatial resolution ($5~\mu$m). This is of major importance to ensure that both qualitative and quantitative measurements can be performed with high reliability on the silver-coated substrates.

\section{Conclusion}
In conclusion, we showed that coating standard microscope slides with a nanoroughened mirror silver film provides high fluorescence signal enhancements. We applied these substrates with tailored roughness to DNA microarrays. Without changing the standard DNA chip reading device, we achieved signal amplifications by more than 40-fold as compared to standard glass slide. We showed that this enhancement is homogeneous at the scale of the scanner resolution, the range of enhancement is broader and the photostability is increased. Therefore, these substrates are of particular interest in bio-sensing applications. Reaching a highly sensitive detection in biosensing allows operating with less biological material, which is of major interest when limited quantities are available, e.g. in biomedical diagnosis. When a 40-fold enhancement meets one's needs the proposed technique permits one to avoid expensive and time-consuming biochemical amplification steps.

\ack
This work was supported by the French National Research Agency through project Plasmoslide. The authors acknowledge Diane Leclerre, Fabrice Richard and Luc Talini of the Genescore Company for DNA microarray applications, and Jean-Marie Herry at UBHM laboratory (INRA, Massy, France) for his precious help with AFM observations.

\section*{References}
\bibliographystyle{unsrt}

\end{document}